\documentstyle[aps,twocolumn,psfig]{revtex}
\begin{document}
\twocolumn[\hsize\textwidth\columnwidth\hsize\csname @twocolumnfalse\endcsname
\title{Rare events and breakdown of 
simple scaling in the Abelian sandpile}
\author{M. De Menech$^{1}$, A. L. Stella$^{1,2}$
and C. Tebaldi$^{1}$ 
}
\address{$^1$ INFM-Dipartimento di Fisica, 
Universit\`a di Padova, I-35131 Padova, Italy\\
$^2$ Sezione INFN, Universit\`a di Padova, I-35131 Padova, Italy 
}
\maketitle

\begin{abstract}
Due to intermittency and conservation, the Abelian sandpile in $2D$ obeys
multifractal, rather than finite size scaling. In the thermodynamic limit, a
vanishingly small fraction of large avalanches dominates the statistics
and a constant gap
scaling is recovered in higher moments of the toppling distribution. Thus,
rare events shape most of the scaling pattern and preserve a meaning for
effective exponents, which can be determined on the basis of numerical and
exact results.

\bgroup\draft
\egroup
\end{abstract}
]

\vspace{1cm}

\narrowtext

The sandpile model was introduced as a first example of self--organized 
criticality (SOC)\cite{BTW}. SOC can manifest itself in
the stationary state of slowly driven, dissipative systems: intermittent
activity bursts occur then over all allowed scales, implying power
law correlations in both space and time.

The analytic tractability of the Abelian sandpile (ASM) has 
allowed a
number of exact results, especially in $2D$
\cite{Dhar,Majumdar,Priezzhevh}. This considerably enhanced the
theoretical interest of this prototype model. On the other hand, in spite of
some remarkable advances \cite{Dhar,Majumdar,PVZ,Ivashkevich}, 
a satisfactory understanding of ASM scaling is still missing. 
Unlike many stationary state properties, the exponents
describing avalanche size distributions are not known exactly, yet,
and remain a major challenge in the whole field of non--equilibrium
critical dynamics. 

By analyzing 
the avalanches of a $2D$ sandpile in terms of waves\cite
{Ivashkevich}, Priezzhev et al.\cite{Priezzhev} proposed a scaling picture
which led to conjecture all exponents, including those describing the
avalanche distributions in terms of covered area and toppling number. 
Unfortunately, determinations of such exponents
are notoriously difficult and there is no compelling numerical support of
the conjecture\cite{Manna1,Lubeck}. 
A more important objection came from an extensive 
study\cite{Paczusky}, which showed the illegitimacy of a basic
assumption made in Ref. 7, concerning the relative sizes and
scaling of successive waves.

In this paper we reconsider scaling in the $2D$ ASM. We show that
intermittency and conservation lead to a breakdown of finite size scaling
(FSS) in the distribution of topplings. This distribution has
manifest multiscaling properties. However, due to a very peculiar role
played by a class of rare, large avalanches, a standard FSS description can be
effectively recovered, if one focuses on the higher moments of the
distribution. Quite remarkably, within our picture the effective scaling
exponents  of the toppling number distribution are determined by an 
asymptotically vanishing fraction of all
avalanches and take the same values conjectured in Ref. 7. Beyond these
results, our analysis exemplifies a novel path towards the correct
characterization of criticality which should be followed also for
other models and phenomena out of equilibrium.

Attempts to identify possible multiscaling features in avalanche
distributions started soon after the introduction of
SOC\cite{Kadanoff}. 
For $1D$ sandpiles evidence has been most often definitely
in favor of multiscaling\cite{Kadanoff,biscaling,Krug}. However, the physical
origin of such multiscaling was never fully elucidated.
On the other hand, for systems in $d>1$
standard FSS has never been put into serious doubt, so far, 
and, with almost no exception \cite{Kadanoff}, has always 
been assumed in both numerical 
and theoretical work. 

We consider a square lattice box of side $L$. At each site $i$ an integer
variable $z_i=1,2,\ldots$ represents the number of grains. If $z_i>z_c=4$ the
site topples, i.e. $z_j\to z_j-\Delta _{ji}$, where $\Delta$ is the
discrete Laplacian. Thus, when toppling, site $i$
loses four grains, giving one of them to each of its neighbors. At the
border, dissipative Dirichlet conditions are assumed, so that one grain
per toppling
leaves the sandpile. Given a stable configuration 
($z_k\leq z_c$ $\forall$ $k$),
the dynamics starts by adding a grain at a randomly chosen site $l$ ($z_l\to
z_l+1$). If $z_l+1>z_c$ the site topples, and, if the case, other sites
topple, as a consequence of this first instability, until a new stable
configuration is reached. The sequence of all such topplings constitutes an
avalanche.

Grain addition is a slow driving mechanism compensated by border
dissipation, which allows to reach the stationary SOC state after
sufficiently many additions. If one considers a large number of
bursts at stationarity, on average, one grain per event must be
dissipated, because each avalanche originates from addition of  one grain.
However, this is still poor information on the dissipation mechanism.
Indeed, outflow of grains occurs intermittently, concentrating on a
small fraction of avalanches, which are separated by random sequences of
non--dissipative bursts. This fraction approaches zero for $L\to\infty$.
Thus, dissipating avalanches increase indefinitely with $L$
their outflow. This fact has crucial
consequences on the role played by this subset of avalanches on the global
statistics.

For the distribution $P$ of the total number of topplings in an
avalanche, $s$, FSS would imply  an asymptotic form 
$P(s,L)\simeq s^{-\tau }p(s/L^D)$, for
$s,L\to\infty$, where $D$ is a
capacity 
fractal dimension 
of the topplings. In order to better identify the role of rare, dissipating
avalanches, we sampled only avalanches generated by
grains added right at the center of the $L\times L$ box.
In this way avalanches
which dissipate a number of grains $n>0$ at the borders are always
necessarily large, since they span a distance of order $L$. Such central
seed setup is also most suitable for our study of correlation functions. 
However as discussed below, our results can be obtained 
also with random additions
over the whole box. Between different central seed additions we decorrelated 
the system with a sufficiently large number of arbitrary avalanches. 

We determined the outflow probability distribution $O(n,L)$ for $%
L=32, 64, 128, 256$ and its moments $\sum_nO(n,L)n^q=\frac{1}{N}\sum_i n_i^q=
<n^q>_L$, where $N$ is the number of sampled avalanches ($ \le 10^8$)
and $n_i$ is the outflow of the $i$-th one.  
Also the
inverse first return frequency, i.e. the average number of avalanches, $T$,
between two successive nonzero outflows was determined.  
The intermittent character of the stationary state
is best revealed by the law $T(L)\propto L^\zeta $, which is satisfied
with $\zeta =0.50\pm 0.01$. Intermittency and stationarity determine a
peculiar scaling of the moments of $O$. Indeed, the fraction of avalanches
with $n>0$ is of order $L^{-\zeta }$. Thus, $\lim_{q\to
0}<n^q>_L\propto L^{-\zeta }$. On the other hand, grain conservation
at stationarity also requires $<n>_L=1=L^0$ (one grain per
avalanche added). Thus, by defining in general $<n^q>_L\propto L^{-\sigma
_n(q)}$, one expects $\sigma _n>0$ ($\sigma_n<0$) for
$0<q<1$ ($q>1$). We verified numerically that the average
outflow, restricted to avalanches with $n>0$, $<n>_{1L}
=\frac{1}{N_1}\sum_{n_i>0} n_i$, 
($N_1$ is the number of avalanches with  
$n>0$), grows as $T$, i.e. $\propto
L^\zeta $. These avalanches rarefy and grow in size
for $L\to\infty$. The most
simple scaling to expect is $<n^q>_L\simeq L^{-\zeta}<n^q>_{1L}
\propto L^{\zeta (q-1)}$ for $q>0$. 
This implies a constant gap $\sigma _n(q)-\sigma
_n(q-1)=-\zeta $ for  $q>1$, i.e. a linear behavior for $\sigma _n$,
consistent with FSS. Such a simple picture is confirmed by our
numerical results. The behavior of $\sigma_n$ is rather close to linear
for all $q>0$ with a gap $\simeq -1/2$ and $\sigma_n(0)\simeq 1/2$.
Some previous determinations of $\zeta$ in the literature fully agree with
the present one\cite{Kadanoff,Manna} and we conjecture $\zeta=1/2$ exactly. 

In order to elucidate the role of outflowing avalanches in
determining the distribution $P$ of the number of topplings, we define $%
g_i({\bf r},L)$ as the number of topplings induced at site $\bf r$ during 
the $i$-th
avalanche. In view of the toppling rules, we clearly have: 
\begin{equation}
\label{eq:nout}
n_i=\sum_{\bf r}{\Delta }g_i({\bf r},L) 
\end{equation}
where the Laplacian acts on $\bf r$ and the sum extends to the whole box. On 
the other hand, $\sum_{\bf r}g_i({\bf r},L)=s_i$ is the number of topplings in
avalanche $i$. Thus, dimensional analysis alone would suggest: 
\begin{equation}
\label{eq:size_and_nout}
<s^q>_L\propto L^{2q}<n^q>_L 
\end{equation}
i.e. $\sigma_s(q)=-2q+\sigma_n(q)$. However, in spite of the strict
linearity of $\sigma_n$ assumed above, a constant gap behavior for $%
\sigma_s$, as suggested by Eq.~\ref{eq:size_and_nout}, 
is not acceptable. First of all, since $s>0$ is not selective 
of outflowing avalanches, we can not have
$\lim_{q\to 0} <s^q>_L\propto L^{-\zeta }$. Thus 
$\sigma_s(0^+)\ne\sigma _n(0^+)$.
Indeed, we know that avalanches with $s>0$ are a nonzero fraction of the
total sampled for $L\to\infty $\cite{Priezzhevh}. This implies 
$\lim_{q\to 0}\sigma_s(q)=0$. In addition, 
$<s>_L=\sum_{\bf r}<g({\bf r},L)>_L \propto L^2$ has
been rigorously shown on the basis of stationarity and
conservation\cite{Dhar}. Thus, while Eq.~\ref{eq:size_and_nout}
 suggests $\sigma_s(q+1)-\sigma
_s(q)=-2.5$ for all $q>0$, basic properties of the non-equilibrium
stationary state impose $\sigma_s(1)-\sigma_s(0)=-2$ exactly.
It remains to be decided whether a gap $-2.5$ still applies to at least
part of the moments of $P$. Below we produce strong evidence
in support of such a conclusion.

\begin{figure}
\centerline{\psfig{figure=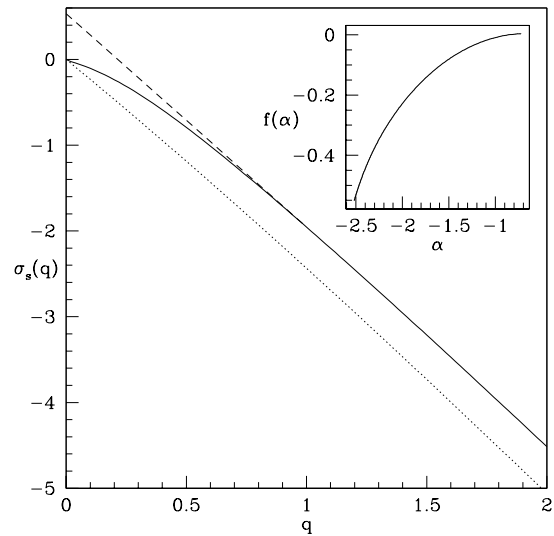,width=3.0in}}
\caption{Plots of $\sigma_s$ (solid line)  
and $\sigma_{1s}$ (dotted line); the dashed line  has
slope $-2.50\pm 0.05$ and y-intercept $\zeta=0.50\pm 0.01$. 
Inset: plot of $f(\alpha)$ {\it vs.} $\alpha$   
extrapolated from avalanche size distributions. 
Assuming $f(\alpha(\infty))=-1/2$
we estimate $\alpha(\infty)=-2.5\pm0.1$.}
\label{fig:sigma}
\end{figure}

Based on  $L=64, 128, 256, 512$, we extrapolated $\sigma_s$ as
reported in Fig.~\ref{fig:sigma}. 
We find $\lim_{q\to0}\sigma_s(q)=0$, and, 
within an accuracy of $10^{-3}$, $\sigma_s(1)=-2$,
as expected. Most remarkably, moments of order $q>1$ appear to conform 
very accurately to the
constant gap $-2.5 \pm 0.1$ expected on the basis of Eq.~\ref{eq:size_and_nout}
(e.g., $\sigma_s(2)=4.5\pm0.05$). Moreover, for $0<q<1$, 
$\sigma_s$ behaves nonlinearly, clearly confirming multiscaling
for $P$. 
Outflowing avalanches alone  obey simple 
scaling to  very high accuracy. 
In Fig.~\ref{fig:sigma} we report the moment exponent $\sigma_{1s}(q)$
for the distribution $P_1$ of $n>0$ avalanches.
Their 
dominance in the cumulative statistics of all avalanches is such that they
impose their constant gap $-2.5$ as soon as $q>1$. 

An alternative way to discuss multiscaling is through the spectrum of
singularity strengths $f(\alpha)$, i.e. the Legendre transform of
$\sigma_s$\cite{Multi}: $\alpha=d\sigma_s/dq$, $f(\alpha)=
-\sigma_s+q\alpha$. $\alpha(q)$ and $f$ are defined with reference to
a saddle point ($s=s^*(q)$) evaluation of $<s^q>_L$ in the $L\to\infty$
limit ($s^*(q)\simeq L^{-\alpha}$, $P(s^*,L)\simeq L^{\alpha+f(\alpha)}$).
The inset in Fig.~\ref{fig:sigma}
 shows the behavior of $f$ as derived by directly
transforming our $L=\infty$ extrapolated data for $\sigma_s$. 
Alternatively one can extrapolate ensemble evaluations
of $f$ and $\alpha$ at finite $L$. The discrepancy of such different
determinations allowed us to estimate their accuracy. 
The shape of $f(\alpha)$ is of course 
consistent with the above results
for $\sigma_s$: in particular, the fact that $f(-5/2)=-1/2$ is very well
satisfied,
agrees with the constant gap $-2.5$ expected for $\sigma_s$ and with the
fact that the linear continuation of the asymptotic
$\sigma_s$ curve (dashed in Fig.~\ref{fig:sigma}) 
intercepts the vertical axis at $y=\zeta=1/2$. 

We verified that the above properties of $P$ and $\sigma_s$ remain
substantially unaltered if sampled avalanches are created by random
grain additions, occurring with uniform probability on the whole box.
In particular, $n>0$ is still a condition able to collect the
subset of large avalanches dominating for $q>1$ in the $L\to\infty$ limit.
Of course, in this case the subset includes also many bursts
of activity (e.g., boundary avalanches), whose importance becomes
negligible in the thermodynamic limit. The central seed scheme offers
the relevant advantage of more rapid convergence with comparable
sizes and statistics. 

The constant gap for $q>1$
and the behavior of $f(\alpha)$ can be fully understood on the 
basis of the dominance
of rare outflowing avalanches. In our sampling, such avalanches are also the
largest, as far as distance spanned is concerned. We can write 
\begin{eqnarray}
\label{eq:gq}
<(\sum_{\bf r} g({\bf r},L))^q>_L=\frac{N_0}{N} <(\sum_{\bf r} 
g({\bf r},L))^q>_{0L} \\ \nonumber
+ \frac{N_1}{N} 
<(\sum_{\bf r} g({\bf r},L))^q>_{1L}
\end{eqnarray}
where $N_0$ is the number of not dissipating
avalanches, and the subscripts of averages indicate restriction to the
corresponding sets. In Eq.~\ref{eq:gq}, of course, $\frac{N_1}N%
\simeq L^{-\zeta }$, and $1-\frac{N_0}N$ $\simeq L^{-\zeta}$.
\begin{figure}
\centerline{\psfig{figure=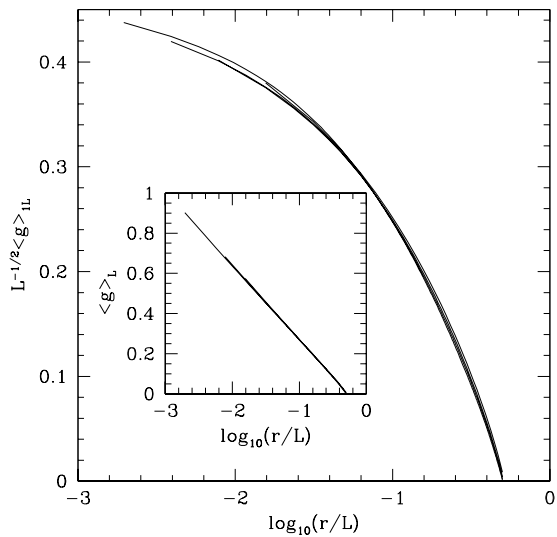,width=3.0in}}
\caption{Collapse fit for 
$<g({\bf r},L)>_{1L}$. The inset reports a 
similar collapse for $<g({\bf r},L)>_L$.
Data refer to $L=64, 128, 256, 512$.}
\label{fig:gcollapse}
\end{figure}

For $q=1$, $<g({\bf r},L)>_L$ and $%
<g({\bf r},L)>_{1L}$  in Eq.~\ref{eq:gq} satisfy very well FSS, 
but in different
forms. 
Our data for $<g({\bf r},L)>_L$ are consistent with a nearly logarithmic
dependence on $r/L$ (up to effects due to the upper and lower cutoffs). 
This is illustrated by the collapse plot in the inset of 
Fig.~\ref{fig:gcollapse}. 
Such 
logarithmic behavior of $<g>_L$ agrees with the fact that
this function has to coincide with the inverse Laplacian  on the
box.
Indeed, due to the local conservation of grains, which holds on average for
the whole sample of avalanches\cite{Dhar}, $<g>_L$ must satisfy
a Poisson equation with a unit source in the origin ${\bf r}=\bf 0$.
On the other hand,
for $<g({\bf r},L)>_{1L}$ one can not invoke local conservation.
In fact, as shown by the collapse plot in Fig.~\ref{fig:gcollapse}, 
our data are very well
consistent
with a form $<g({\bf r},L)>_{1L}\simeq L^{1/2}g_1({\bf r}/L)$, and the function
$g_1$ is clearly not logarithmic. The peculiar anomalous scaling 
dimension $1/2$
appearing in $<g>_{1L}$ should be identified with $\zeta$. 
The collapses in Fig.~\ref{fig:gcollapse} show that $n>0$ avalanches are
indeed much richer in topplings than the global average.
The average number of waves in each outflowing
avalanche, coinciding with $<g({\bf 0},L)>_{1L}$\cite{Majumdar,Ivashkevich}, 
is proportional to $L^{1/2}$. In fact, if referred 
to all avalanches, such an average number grows only
logarithmically with $L$\cite{Majumdar}, as we verified to high accuracy.
Both terms on the r.h.s. of Eq.~\ref{eq:gq} give a
contribution proportional to $L^2$ when $L\to\infty$, as expected for the
l.h.s.. 
    
The singularity
strength $f(\alpha)$ is also fully consistent
with the properties discussed above. $f(-2.5)=-1/2$ indicates that outflowing
avalanches, which are a fraction $L^{-1/2}$ of the total, have a fractal 
dimension
$D=2.5$. In fact, within the multiscaling framework one obtains a whole
continuum of fractal dimensions
$D_q=[\sigma_s(1)-\sigma_s(q)]/[q-1]$ 
for avalanches.  
$D_q$\cite{Multi} 
ranges in the whole interval
$(2,2.5)$, for $0\leq q\leq 1$, with $D_0=2$ and 
$D_1=2.5$ ($D_q=D_1$ for $q>1$).

Clearly, the very notion of standard scaling exponents can not
adapt to $P$, due to the multiscaling concentrated at low $q$.
However, in numerical work, $P$ is usually analyzed by assuming a FSS form,
e.g. on the basis of collapse fits.
Thus, it is important to ask what should be the FSS form of
distribution which most accurately reproduces the scaling of the moments of 
$P$. 
The optimal job in this respect is done by
a distribution of the FSS form $s^{-\tau}p(s/L^D)$, whose 
$q$-th moment exponent
coincides with $\sigma_s(q)$ for $q>1$, and with the function
represented by the straight dashed line in Fig.~\ref{fig:sigma} for $q<1$.
Of course, we put the fractal dimension equal to $D=D_1=2.5$ in
$p$, because its opposite has to coincide with 
the asymptotic slope of $\sigma_s$.
For such a scaling function one gets
$\sum_s s^{1-\tau}p(s/L^D)\simeq L^{D(2-\tau)}$.
Thus, $D(2-\tau )=2$ leads to
$\tau =6/5$. Some accurate numerical
determinations  based on FSS\cite{Manna1} are very close to this $\tau$, 
which remarkably coincides with
that conjectured in Ref. 7.
The geometry of the various lines
in Fig.~\ref{fig:sigma} is very eloquent. In order to reduce the
multiscaling of $P$ to simple FSS, we have to shift to $s=0$
the contribution of 
$n=0$ avalanches
in the histogram of $P$. This amounts to bend the 
actual $\sigma_s$ curve in the interval
$(0,1)$ into a straight segment (dashed), with intercept $\zeta$ 
on the vertical axis.
Our results for $D_q$  and $\tau$
give an indication of the problems arising when, e.g., one tries
to enforce FSS collapses of data for the simultaneous determination
of $D$ and $\tau$. Privileging collapse in the region of low $s/L^D$
produces a simultaneous lowering of both $D$ and $\tau$, like
when searching a solution of $D_q(2-\tau)=2$  for low values
of $q$. The extreme case is $\tau =1 $ with $D=D_0=2$. 
Emphasizing collapse at high $s/L^D$ produces higher determinations
of both $D$ and $\tau$. In this case a relatively poor sampling
can lead to an overestimation of both exponents with respect to the
above effective values. Indeed, we verified that 
$\sigma_s$  and thus $D_q$  tend to be overestimated 
systematically at large $q$.

The standard
exponents of $n>0$  avalanches alone (ensemble 1) are $\tau _1$ and 
$D$ such
that we can put $P_1(s,L)=s^{-\tau _1}p_1(s/L^D)$. From our results and from
the constant gap $-2.5$ for $\sigma _{1s}$, it follows
immediately $D=2.5$ and $\tau _1=1$.

Summarizing, we showed that standard FSS does not hold for the
ASM in $2D$. Conservation of grains is guaranteed by an intermittent
mechanism, with rare, large avalanches producing the outflowing
current. These avalanches are such to fully determine
the moments of $P$ from the first one upwards. The
multiscaling spectrum of singularity strengths for P has peculiar features
associated with the dominance of rare events. In particular 
$f(-5/2)=-1/2$ shows that rare avalanches have a fractal dimension
equal to $5/2$ and occur with frequency $\propto L^{-1/2}$. 
These features are such to give $\tau=6/5$ and $D=2.5$ as
effective exponents representing the multiscaling within an imposed
simple FSS framework. Not surprisingly, FSS based
numerical determinations of $\tau$
and $D$ are often quite close to the values mentioned above \cite{Manna1}.

In spite of
the fact that the approach of Ref. 7 does not take 
into account multiscaling features, 
the values of $\tau$ and $D$ we propose  
coincide with those conjectured there.
Not surprisingly, in view of our results, that conjecture has met
problems of numerical verification\cite{Paczusky}. A
revisitation of wave, or, rather, cluster \cite{condmat} properties 
in the light of the dominance of large avalanches and multiscaling can 
clarify the situation\cite{preparation2}. 

Crucial
to the identification of our effective 
exponents are the constant gap $-(2+\zeta)$ for 
$q>1$ and the exact result $\sigma_s(1)=-2$\cite{Dhar}. Thus, it is
precisely $\zeta=1/2$ that determines the conjectured $\tau$ and $D$ in
the present framework.
The critical state of the ASM in $2D$ is expected to correspond
to that of a non-unitary conformal field theory with central charge
$c=-2$\cite{Majumdar}. The compatibility
of multiscaling with other non-unitary conformal field theories has
been pointed out recently\cite{Mudry}. It is also known\cite{Saleur} that in
a theory with $c=-2$ a correlator of disorder operators possesses
exactly the scaling  dimension $1/2$.

We expect intermittency and
dominance of rare, large events to play a key role 
also in the physics of other SOC models. Indeed, 
$1D$ sandpiles most often
display pronounced multiscaling features and
intermittency \cite{Kadanoff,biscaling,Krug}. 
Of course, the Abelian symmetry and the
Laplacian characterization of the toppling dynamics made the analysis
of these features relatively easy here, to the extent that exact
properties of the model could be inferred. 
The case of ASM in higher dimensions, 
for which $\sum_{\bf r}<g({\bf r},L)>_L\propto L^2$ seems to 
hold irrespective
of $d$\cite{Dhar,Chessa}, is also quite intriguing.

Partial support from the European Network Contract ERBFMRXCT980183 is 
acknowledged.
C.T. wishes to thank ISAS, Trieste, where part of 
this work was done.~\footnote{demenech@pd.infn.it, 
stella@pd.infn.it, tebaldi@pd.infn.it.}

\end{document}